\documentclass[reprint, superscriptaddress, secnumarabic, amssymb, nobibnotes, aps, prl]{revtex4-1}

\usepackage{graphicx}
\usepackage{epstopdf}
\usepackage[T1]{fontenc}
\usepackage[latin9]{inputenc}
\usepackage{amsbsy}
\usepackage{gensymb}
\usepackage[T1]{fontenc}
\usepackage[latin9]{inputenc}
\usepackage{amsmath}
\usepackage{amssymb}
\usepackage{bbm}
\usepackage{braket}
\usepackage{xcolor}
\allowdisplaybreaks
\usepackage{graphicx}
\usepackage[colorlinks=true]{hyperref}  
\hypersetup{
    bookmarks=true,         
    unicode=false,          
    pdftoolbar=true,        
    pdfmenubar=true,        
    pdffitwindow=false,     
    pdfstartview={FitH},    
    pdftitle={Muon},    
    pdfauthor={R. P. Singh},     
    pdfsubject={},   
    pdfcreator={},   
    pdfproducer={}, 
    pdfkeywords={} {} {}, 
    pdfnewwindow=true,      
    colorlinks=true,       
    linkcolor=blue, 
    citecolor=blue,        
    filecolor=magenta,      
    urlcolor=blue           
} 
\usepackage[normalem]{ulem}

\newcommand{\equref}[1]{Eq.~(\ref{#1})}

\newcommand{\figref}[1]{Fig.~\ref{#1}}

\renewcommand{\approx}{\simeq}

\begin{document}
\title{\textrm{Superconductivity in equimolar Nb-Re-Hf-Zr-Ti high entropy alloy}}
\author{Sourav Marik}
\affiliation{Indian Institute of Science Education and Research Bhopal, Bhopal, 462066, India}
\author{Maneesha Varghese}
\affiliation{Indian Institute of Science Education and Research Bhopal, Bhopal, 462066, India}
\author{K. P. Sajilesh}
\affiliation{Indian Institute of Science Education and Research Bhopal, Bhopal, 462066, India}
\author{Deepak Singh}
\affiliation{Indian Institute of Science Education and Research Bhopal, Bhopal, 462066, India}
\author{R. P. Singh}
\email[]{rpsingh@iiserb.ac.in}
\affiliation{Indian Institute of Science Education and Research Bhopal, Bhopal, 462066, India}

\date{\today}
\begin{abstract}
\begin{flushleft}

\end{flushleft}
Herein, we report the synthesis and detailed superconducting properties of a new high entropy alloy superconductor with nominal composition Nb$_{20}$Re$_{20}$Zr$_{20}$Hf$_{20}$Ti$_{20}$ using powder X-ray diffraction (XRD), energy-dispersive X-ray spectroscopy (EDX), magnetization, transport, and thermodynamic measurements. The room temperature powder XRD confirms that the alloy is arranged on a simple body centered cubic crystal lattice with lattice parameter a = 3.38 (1) $\text{\AA}$. EDX measurement yields an average composition of Nb$_{21}$Re$_{16}$Zr$_{20}$Hf$_{23}$Ti$_{20}$ (in atomic $\%$). Transport, magnetic and heat capacity measurements reveal that the material is a type-II superconductor with the bulk superconducting transition at $T_{c}$ = 5.3 K, lower critical field $H_{c1}$(0) = 33 mT and upper critical field $H_{c2}$(0) = 8.88 T. Low temperature specific heat measurement indicates that the sample is a moderately coupled superconductor, and the electronic specific heat data fits well with the single-gap BCS model. 
\end{abstract}
\maketitle

\section{Introduction}

In recent times, the advent of high entropy alloys (HEAs) has attracted immense interest in the exploration of exciting and tunable mechanical and physical properties and indeed, has initiated a fascinating interdisciplinary research topic \cite{1,2,3,4,5}. This new class of materials is of particular interest due to their potential for the development of novel materials with combinations of properties superior to those of conventional alloys and phases. HEAs are generally constructed from five or more principal elements in equimolar or near equimolar ratio \cite{1}. By applying this novel design approach, several HEAs with simple structures such as a body-centered cubic (bcc) or a face-centered cubic (fcc) structures have been explored \cite{2,3,4,5}. The high entropy of mixing ($\Delta S_{mixing}$) stabilizes disordered solid solution phases in simple crystallographic lattices, and commonly their structures are notified by a topologically ordered lattice with an extremely high chemical disorder.

 Besides the diverse possibility of compositional variety, the HEAs show several novel and tunable functional properties. Illustrative examples include superior mechanical performance at high and cryogenic temperatures \cite{6}, concurrent strength and ductility \cite{7,8}, high mechanical strength, corrosion resistance and thermal stability \cite{2,3,5,9}. In addition to their excellent mechanical properties, exotic electronic properties such as complex magnetism  \cite{6} and superconductivity \cite{11,12,13,14,15} are discovered recently in HEAs. However, to date, among several studied HEAs only two pentanary superconducting materials are discovered. Ta-Nb-Hf-Zr-Ti system, with composition [TaNb]$_{0.67}$[HfZrTi]$_{0.33}$ is the first reported HEA superconductor (bcc cubic structure) with $T_{c}$ $\approx$ 7.3 K, and shows an upper critical field $\mu_{0}$$H_{c2}$(0) = 8.2 T \cite{11,12,13,14}. Quite remarkably, this compound shows an extraordinary robust zero-resistance superconductivity at pressure up to 190 GPa \cite{16}. Recently, another pentanary HEA system with composition [ScZrNb]$_{0.65}$[RhPd]$_{0.35}$ (cubic primitive CsCl-type lattice) is reported to show superconductivity with $T_{c}$ $\sim$ 9.3 K, and exhibits $\mu_{0}$$H_{c2}$(0) = 10.7 T \cite{15}. It is worth mentioning here that the Ta-Nb-Hf-Zr-Ti HEA superconductor is derived from Nb-Ti-based binary alloys, which are nowadays the widely used materials for superconducting magnets. Therefore, the discovery of bulk superconductivity on a highly disordered lattice in the Ta-Nb-Hf-Zr-Ti HEA and the continuous existence of zero resistance under pressure up to 190 GPa highlight the potentiality of HEA superconductors for future applications including application under extreme conditions. In general, considering the aforesaid superior and improved mechanical and material characteristics, superconducting HEAs show promises to be a material for future superconducting applications. 
 
 In this paper, we present the superconducting properties of equimolar HEA compound Nb$_{20}$Re$_{20}$Zr$_{20}$Hf$_{20}$Ti$_{20}$ exhibiting bulk superconductivity at ($T_{c}$) 5.3 K. This is in fact the first bcc structured equimolar HEA superconductor. Superconducting properties were determined by resistivity, magnetization, and specific heat measurements.
	
\section{Experimental Details}

A polycrystalline sample of nominal composition Nb$_{20}$Re$_{20}$Zr$_{20}$Hf$_{20}$Ti$_{20}$ (2g) was synthesized using the standard arc-melting technique. Stoichiometric amounts of Nb (purity 99.8 $\%$), Re (purity 99.99 $\%$), Zr (purity 99.95 $\%$), Hf (purity 99.7 $\%$) and Ti (purity 99.99 $\%$) pieces were taken on a water cooled copper hearth under the flow of high purity argon gas. They were melted in high current (T>2500) to make a single button, then flipped and re-melted several times for the sample homogeneity. The sample formed was hard with insignificant weight loss. Powder X-ray diffraction (XRD) was carried out at room temperature (RT) on a PANalytical diffractometer equipped with Cu-K$_{\alpha}$ radiation ($\lambda$ = 1.54056 \text{\AA}). Sample compositions were checked by a scanning electron microscope (SEM) equipped with an energy-dispersive X-ray (EDX) spectrometer. DC magnetization and ac susceptibility measurements were performed using a Quantum Design superconducting quantum interference device (MPMS 3, Quantum Design). The electrical resistivity measurements were performed on the physical property measurement system (PPMS, Quantum Design, Inc.) by using a conventional four-probe ac technique at frequency 17 Hz and excitation current 10 mA. The measurements were carried out under the presence of different magnetic fields. Specific heat measurement was performed by the two tau time-relaxation method using the PPMS in zero magnetic field.

\section{Results and Discussion}

\begin{figure}
\includegraphics[width=1.0\columnwidth]{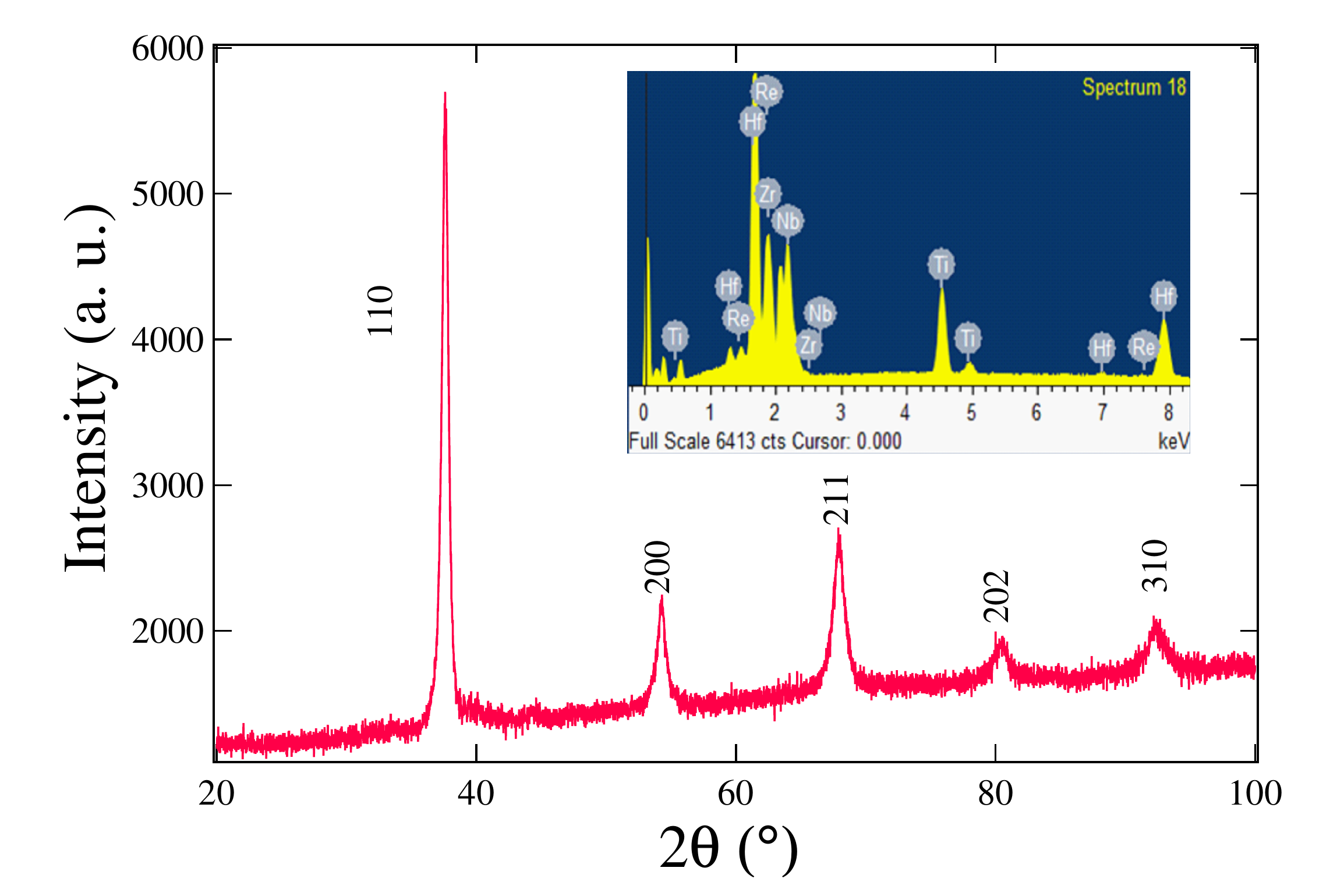}
\caption{\label{Fig1:XRD} Room temperature (RT) powder XRD pattern for Nb$_{21}$Re$_{16}$Zr$_{20}$Hf$_{23}$Ti$_{20}$. Inset shows the RT EDX pattern of the material.}
\end{figure}

\begin{figure}
\includegraphics[width=1.0\columnwidth]{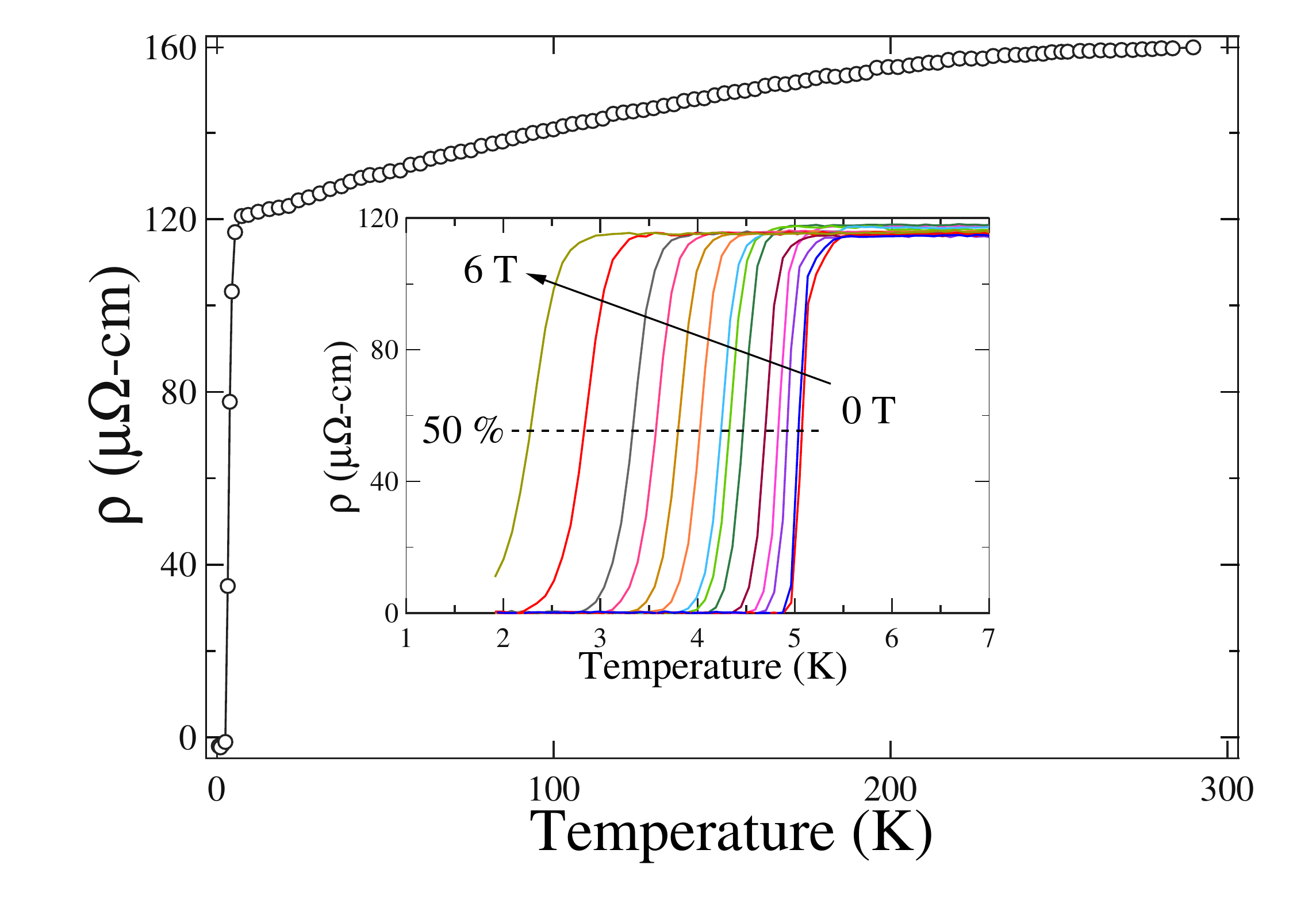}
\caption{\label{Fig2:RESISNO} (a) The resistivity measurement $\rho(T)$ for Nb$_{21}$Re$_{16}$Zr$_{20}$Hf$_{23}$Ti$_{20}$ taken in zero field in a temperature range of 1.85 K $\le$ T $\le$ 290 K. Inset shows the magnetic field variation (up to 6 T) of resistivity for the same sample.}
\end{figure}

\begin{figure}
\includegraphics[width=1.0\columnwidth]{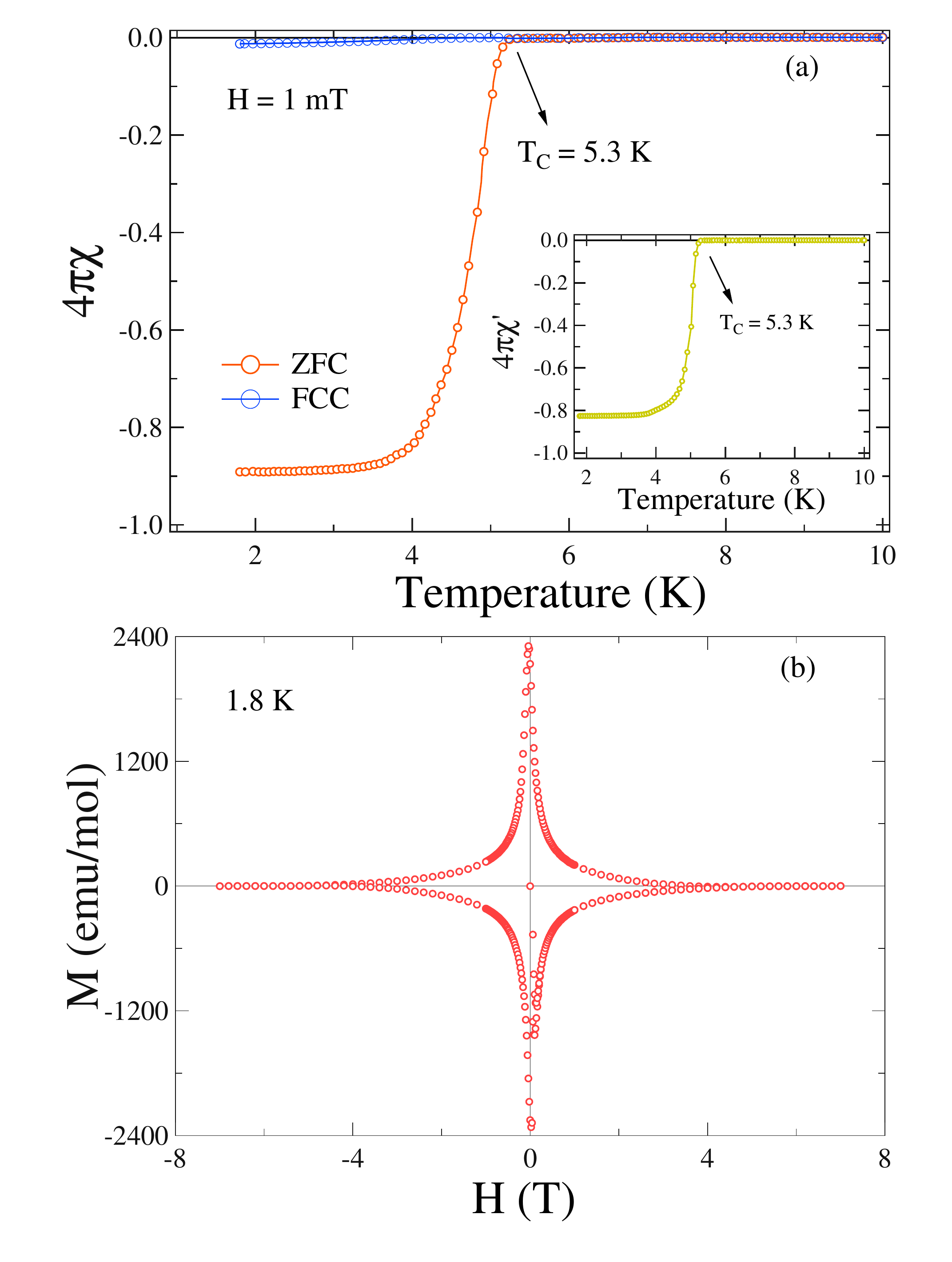}
\caption{\label{Fig3:ZFC} (a) The temperature variation of dc susceptibility for Nb$_{21}$Re$_{16}$Zr$_{20}$Hf$_{23}$Ti$_{20}$ taken in 1 mT field shows the superconducting transition at $T_{c}$ = 5.3 K. Inset shows the ac susceptibility measurement. (b) The magnetization curve at T = 1.8 K, in high applied magnetic field range ($\pm$ 7T).}
\end{figure}

\begin{figure}
\includegraphics[width=1.0\columnwidth]{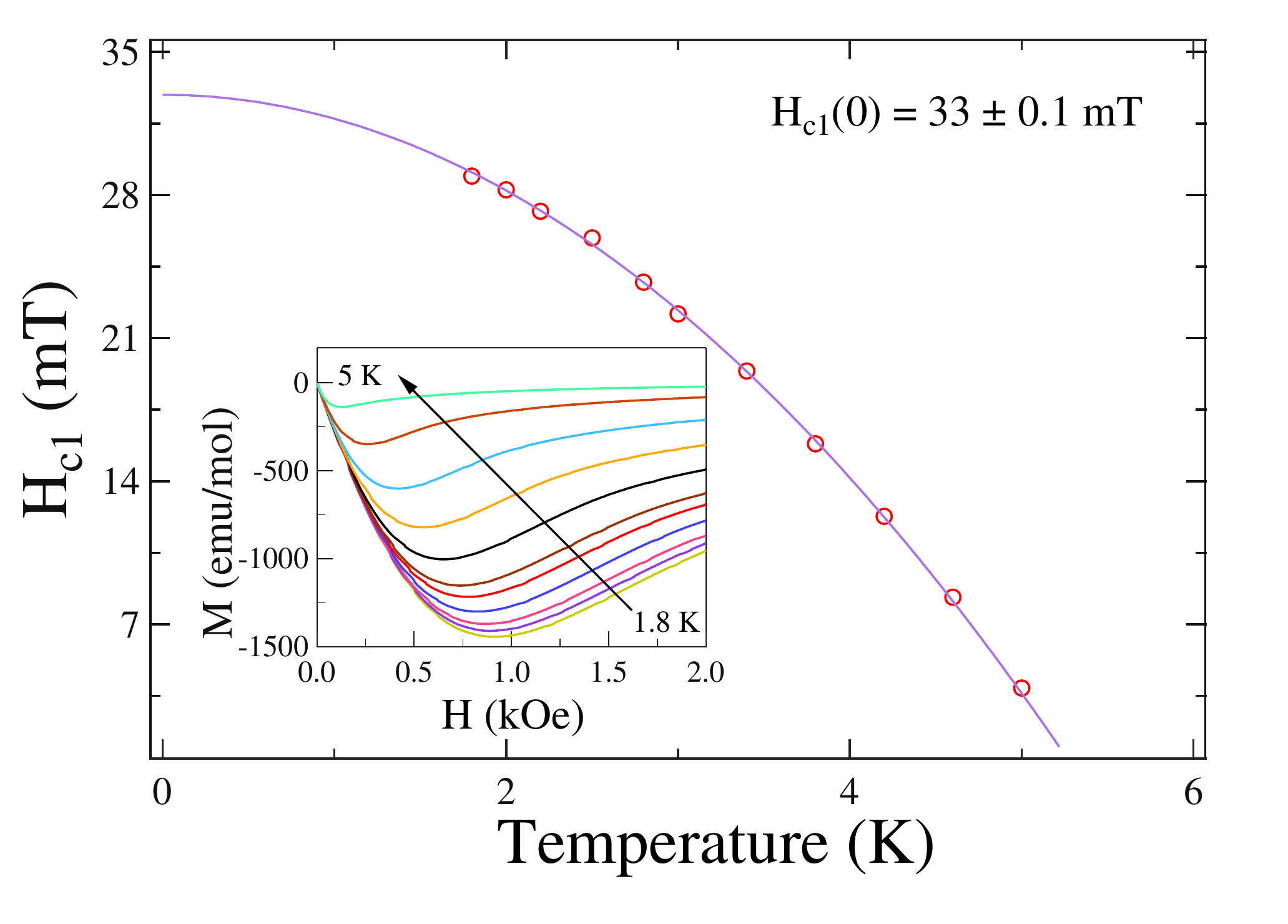}
\caption{\label{Fig4:HC1} (a) Temperature dependence of the lower critical field $H_{c1}$ and the fitting using Ginzburg-Landau relation gives $H_{c1}$(0) = 33 $\pm$ 1 mT. Inset: The magnetization curves as a function of applied magnetic field at various temperatures.}
\end{figure}

\begin{figure}[h]
\includegraphics[width=1.0\columnwidth]{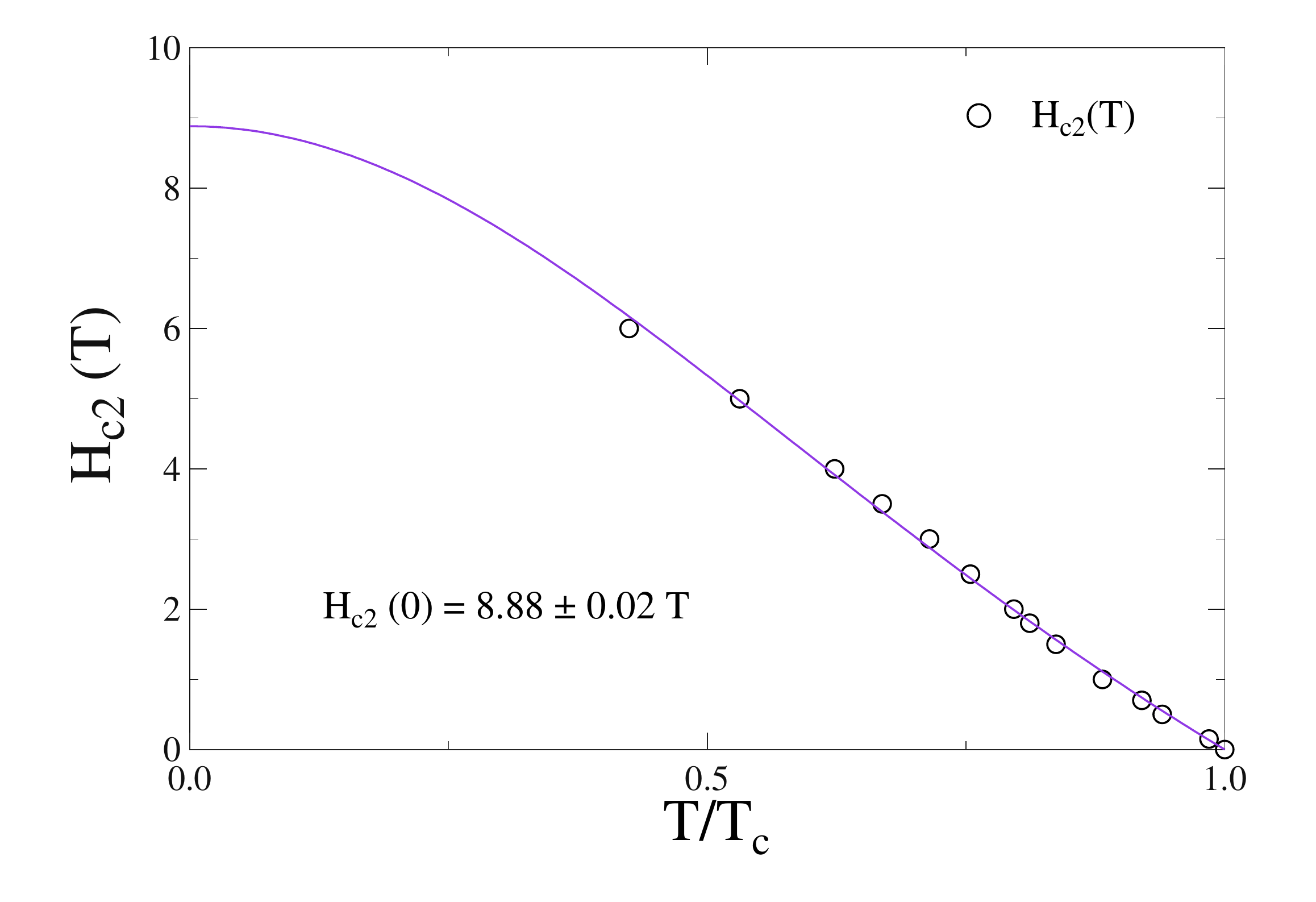}
\caption{\label{Fig5:HC2} The upper critical field $H_{c2}$(T) obtained from resistivity measurements. The dotted line shows the GL fit, yielding $H_{c2}$(0) $\simeq$ 8.88 $\pm$ 0.02 T for  Nb$_{21}$Re$_{16}$Zr$_{20}$Hf$_{23}$Ti$_{20}$.}
\end{figure}

The RT powder XRD pattern (\figref{Fig1:XRD}) indicates that the sample can be isolated as a single phase and can be indexed with a simple bcc unit cell with lattice parameter a = 3.38 (1) \text{\AA}. The broadness in the XRD peaks could be attributed to the high degree of disorder present in this HEA material. Also, the hardness of the sample makes it difficult to prepare fine particle size powder for diffraction experiment. Therefore, the non-ideal sample preparation may have a contribution to the broadening of the XRD peaks. Inset in \figref{Fig1:XRD} shows the RT EDX pattern of the material. The EDX analysis on different part of the sample yields an average composition of Nb$_{21}$Re$_{16}$Zr$_{20}$Hf$_{23}$Ti$_{20}$ (in atomic $\%$) with less than 1 $\%$ uncertainty for each element, and hereafter we have used this composition in the manuscript.\\ 
\figref{Fig2:RESISNO} shows the temperature variation of zero field resistivity, which highlights a sharp drop to a zero resistivity superconducting state below 5.3 K. The normal state resistivity is found to increase leisurely with increasing temperature illustrating a poor metallic behavior. The residual resistivity ratio (RRR) is found to be $\rho$(290)/$\rho$(10) = 1.3, which is very small, suggesting the existence of atomic scale disorder in the sample. The small RRR value is comparable to those of previously reported HEAs \cite{11,14,15}. The inset in \figref{Fig2:RESISNO} shows the magnetic field dependence of resistivity ($\rho$(H)-T), and this shows a gradual suppression of $T_{c}$ with increasing magnetic field.
Fig. \ref{Fig3:ZFC}(a) shows the temperature variation of dc and ac magnetic susceptibility measurements. These confirm the existence of bulk superconductivity in the sample with a superconducting transition at $T_{c}$ = 5.3 K. The measurements were performed between 1.8 K to 8 K, with zero field cooling (ZFC) and field cooling (FC) mode in an external field of 1 mT.
Fig. \ref{Fig3:ZFC}(b) shows the magnetic field variation of magnetization (M-H) curve at 1.8 K (up to $\pm$ 7T). As evident from the graph, the sample exhibits type-II superconductivity. Lower critical field $H_{c1}$(T) (\figref{Fig4:HC1}), which is defined as the field deviating from the linear line for initial slope in magnetization curve is found to be 33 mT at zero temperature in accordance with the Ginzburg-Landau approximation. 
As described previously (\figref{Fig2:RESISNO}), the magnetic field variation of resistivity highlights a gradual suppression of $T_{c}$ with increasing magnetic field. The critical temperatures are derived from the midpoint values of the superconducting transition in the resistivity measurements. The temperature variation of the upper critical field ($H_{c2}$), calculated from ($\rho$(H)-T) measurements is shown in \figref{Fig5:HC2}. The experimental $H_{c2}$ can be described by the Ginzburg-Landau expression. 
\begin{equation}
H_{c2}(T) = H_{c2}(0)\frac{(1-(T/T_{c})^{2})}{(1+(T/T_{c})^2)}
\label{eqn1:hc2}
\end{equation} 

The estimated value of the upper limit for the upper critical field $H_{c2}$(0) is 8.88 T, which is close to the BCS weak coupling Pauli limit ($H^{p}_{c2}$(0) = 1.84$T_{c}$, $T_{c}$ = 5.3 K gives $H^{p}_{c2}$(0) $\approx$ 9.7 T). In comparison, the similar bcc structured [TaNb]$_{0.67}$[HfZrTi]$_{0.33}$ ($T_{c}$ = 7.3 K) shows $H_{c2}$(0) = 8.3 T, which is, in fact far below the BCS weak coupling Pauli limit ($H^{p}_{c2}$(0) $\approx$ 13.4 T) \cite{11}. The enhanced spin orbit coupling due to the introduction of Re in the structure might be playing a role in the enhancement of $H_{c2}$(0) for the present sample. However, further experiments are required to establish a relationship between $H_{c2}$ and the magnitude of the spin orbit coupling. The Ginzburg-Landau coherence lengths $\xi_{GL}$(0), estimated from the following relation \cite{17} is 6.1 nm.
 
\begin{equation}
H_{c2}(0) = \frac{\Phi_{0}}{2\pi\xi_{GL}^{2}}
\label{eqn2:up}
\end{equation}

\begin{figure}
\includegraphics[width=1.0\columnwidth]{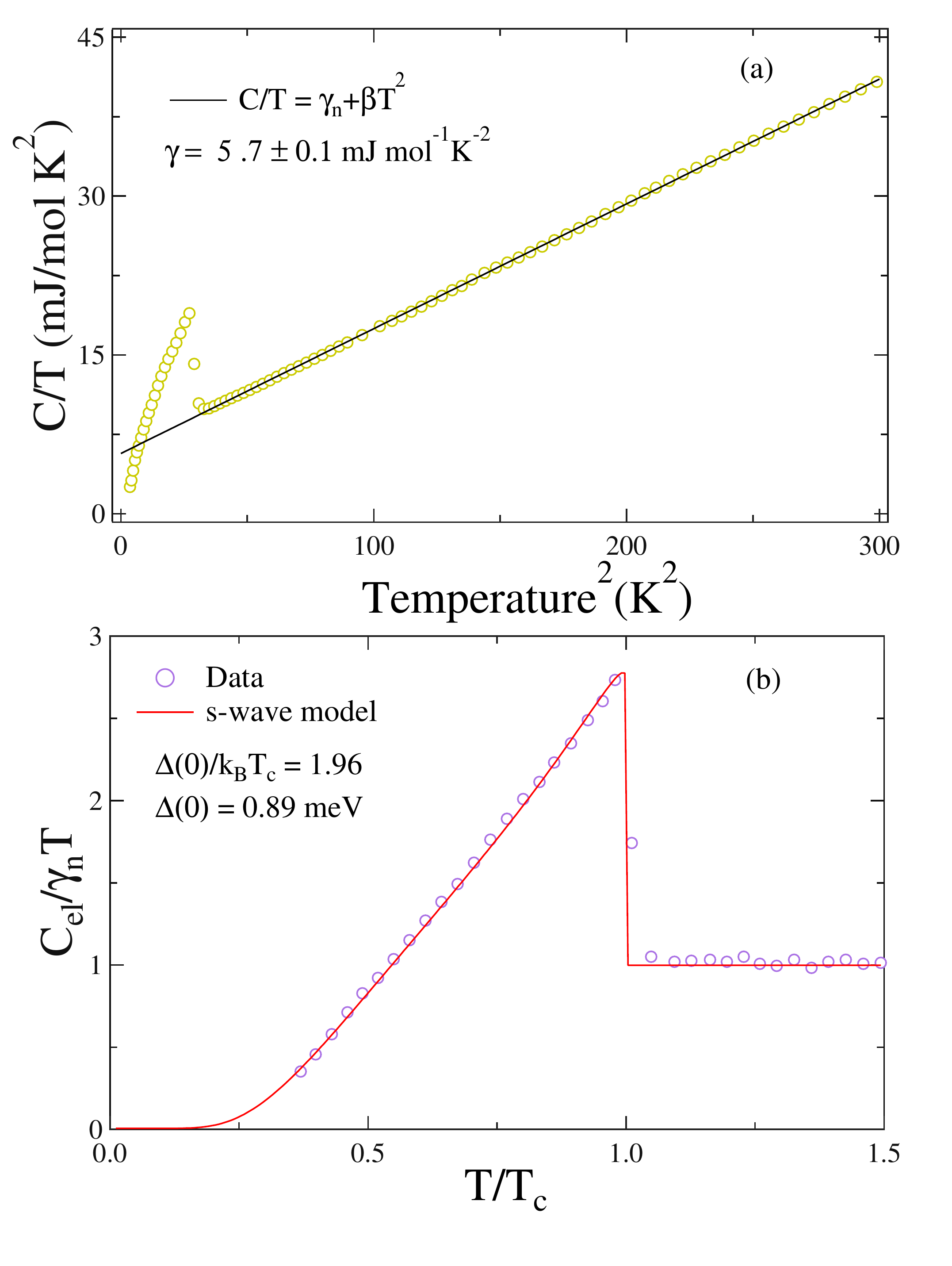}
\caption{\label{Fig6:C/T} (a) The C/T vs T$^{2}$ data in the temperature range 3 K $\le$ $\textit{T}$ $\le$ 100 K, fitted with low temperature Debye model  $C_{P}/T$ = $\gamma$ + $\beta$$T^{2}$. (b) The low temperature specific heat data in the superconducting regime fitted with the single gap s-wave model using \equref{eqn5:Cel}.} 
\end{figure}

To analyze the phonon properties and electronic density of states, we have performed the heat capacity measurements for the Nb$_{21}$Re$_{16}$Zr$_{20}$Hf$_{23}$Ti$_{20}$ sample. Fig. \ref{Fig6:C/T} illustrates the $C_{P}/T$ as a function of $T^{2}$ above 1.8 K. The superconducting transition is observed at 5.3 K (a jump in the heat capacity data) and this confirms the bulk superconductivity in the sample. The fitting of low-temperature normal state specific heat using the equation  $C_{P}/T$ = $\gamma$ + $\beta$$T^{2}$ ($\gamma$ = Sommerfeld coefficient and $\beta$ is the lattice contribution to the specific heat) yields $\gamma$ = 5.7 mJ/mol K$^{2}$ and $\beta$ = 0.1178 mJ/mol K$^{4}$. The Debye temperature $\theta_{D}$ is obtained using the formula $\theta_{D} = \left(12\pi^{4}RN/5\beta\right)^{1/3}$ where R is the molar gas constant (=8.314 J mol$^{-1}$ K$^{-1}$). Using N = the number of atoms per formula unit = 1, it gives $\theta_{D}$ = 254.6 K. The density of states value at Fermi level [$N(E_{F})$ =  $\frac{3\gamma}{\pi^{2}K_{B}^{2}}$] is calculated as 2.42 states eV$^{-1}$ f.u.$^{-1}$ for Nb$_{21}$Re$_{16}$Zr$_{20}$Hf$_{23}$Ti$_{20}$. The electron-phonon coupling constant, which gives the strength of the attraction between the electron and phonon can be calculated using the McMillan formula \cite{18} 

\begin{equation}
\lambda_{e-ph} = \frac{1.04+\mu^{*}ln(\theta_{D}/1.45T_{c})}{(1-0.62\mu^{*})ln(\theta_{D}/1.45T_{c})-1.04 }
\label{eqn3:ld}
\end{equation} 

By considering the Coulomb pseudopotential $\mu^{*}$ = 0.13, commonly used for many intermetallic superconductors \cite{14,19,20}, $\lambda_{e-ph}$ is calculated as 0.69. The normalized specific heat jump ($\frac{\Delta C_{el}}{\gamma_{n}T_{c}}$) is obtained as 1.81 for Nb$_{21}$Re$_{16}$Zr$_{20}$Hf$_{24}$Ti$_{20}$, which is higher than the BCS value of 1.43 in the weak coupling limit and indicating moderately coupled superconductivity in the sample.\\

We have also estimated the superconducting gap using the low temperature specific heat data. According to the BCS theory, the temperature dependence of the specific heat in the superconducting state can best be described by the single-gap BCS expression for normalized entropy S

\begin{equation}
\frac{S}{\gamma_{n}T_{c}} = -\frac{6}{\pi^2}\left(\frac{\Delta(0)}{k_{B}T_{c}}\right)\int_{0}^{\infty}[ \textit{f}\ln(f)+(1-f)\ln(1-f)]dy \\
\label{eqn4:s}
\end{equation}

where $\textit{f}$($\xi$) = [exp($\textit{E}$($\xi$)/$k_{B}T$)+1]$^{-1}$ is the Fermi function, $\textit{E}$($\xi$) = $\sqrt{\xi^{2}+\Delta^{2}(t)}$, $\xi$ = energy of normal electrons measured relative to the Fermi energy, $\textit{y}$ = $\xi/\Delta(0)$, $\mathit{t = T/T_{c}}$, and $\Delta(t)$ = tanh[1.82(1.018(($\mathit{1/t}$)-1))$^{0.51}$] is the BCS approximation for the temperature dependence of the energy gap. 

The normalized electronic specific heat $C_{el}$ below $T_{c}$ can be described by relation

\begin{equation}
\frac{C_{el}}{\gamma_{n}T_{c}} = t\frac{d(S/\gamma_{n}T_{c})}{dt} \\
\label{eqn5:Cel}
\end{equation}

The fitting of specific heat data using \equref{eqn5:Cel} yields $\alpha$ = $\Delta(0)/k_{B}T_{c}$ = 1.96 $\pm$ 0.03. The obtained value of $\alpha$ = $\Delta(0)/k_{B}T_{c}$ is higher than the BCS value $\alpha_{BCS}$ = 1.764 in the weak coupling limit, suggesting a moderately coupled superconductivity in Nb$_{21}$Re$_{16}$Zr$_{20}$Hf$_{23}$Ti$_{20}$. The obtained values of normalized specific heat jump and $\lambda_{e-ph}$ also support the moderately coupled superconducting nature in the sample.\\

To further analyze the nature of the superconductivity in this new material, we have estimated the Fermi temperature (T$_{F}$) of the material.
\begin{equation}
 k_{B}T_{F} = \frac{\hbar^{2}}{2}(3\pi^{2})^{2/3}\frac{n^{2/3}}{m^{*}}, 
\label{eqn6:tf}
\end{equation}
where n is the quasiparticle number density per unit volume. 
Using the following equation and the Sommerfeld coefficient for Nb$_{21}$Re$_{16}$Zr$_{20}$Hf$_{23}$Ti$_{20}$, we can calculate the quasiparticle number density per unit volume and mean free path \cite{ck}
\begin{equation}
\gamma_{n} = \left(\frac{\pi}{3}\right)^{2/3}\frac{k_{B}^{2}m^{*}V_{\mathrm{f.u.}}n^{1/3}}{\hbar^{2}N_{A}}
\label{eqn7:gf}
\end{equation}
where k$_{B}$ is the Boltzmann constant, N$_{A}$ is the Avogadro constant, V$_{\mathrm{f.u.}}$ is the volume of a formula unit and m$^{*}$ is the effective mass of quasiparticles. The electronic mean free path $\textit{l}$ is related to residual resistivity $\rho_{0}$ by the equation
 \begin{equation}
\textit{l} = \frac{3\pi^{2}{\hbar}^{3}}{e^{2}\rho_{0}m^{*2}v_{\mathrm{F}}^{2}}
\label{eqn8:le}
\end{equation}
where the Fermi velocity $v_{\mathrm{F}}$ is related to the effective mass and the carrier density by
\begin{equation}
n = \frac{1}{3\pi^{2}}\left(\frac{m^{*}v_{\mathrm{f}}}{\hbar}\right)^{3} .
\label{eqn9:n}
\end{equation}
In the dirty limit, the penetration depth $\lambda_{GL}$(0) can be estimated by 
\begin{equation}
\lambda_{GL}(0) = \lambda_{L}\left(1+\frac{\xi_{0}}{\textit{l}}\right)^{1/2}
\label{eqn10:f}
\end{equation}
where $\xi_{0}$ = BCS coherence length. The London penetration depth ($\lambda_{L}$) is given by
\begin{equation}
\lambda_{L} = \left(\frac{m^{*}}{\mu_{0}n e^{2}}\right)^{1/2}
\label{eqn11:laml}
\end{equation}
The Ginzburg-Landau coherence length is also affected in the dirty limit. The relationship between the BCS coherence length $\xi_{0}$ and the Ginzburg-Landau coherence $\xi_{GL}$(0) at T = 0 is
\begin{equation}
\frac{\xi_{GL}(0)}{\xi_{0}} = \frac{\pi}{2\sqrt{3}}\left(1+\frac{\xi_{0}}{\textit{l}}\right)^{-1/2}
\label{eqn12:xil}
\end{equation}
Equations (14)-(19) form a system of four equations and can be used to estimate the parameters m$^{*}$, n, $\textit{l}$, and $\xi_{0}$ as done in Ref.\cite{DAM}. These equations were solved simultaneously using the values $\gamma_{n}$ = 5.7 mJ mol$^{-1}$K$^{-2}$, $\xi_{GL}$(0) = 6.1 nm, and $\rho_{0}$ = 120.01 $\mu$$\Omega$-cm. The estimated value of $\xi_{0}$ if found to be larger than $\textit{l}$, indicating that Nb$_{21}$Re$_{16}$Zr$_{20}$Hf$_{23}$Ti$_{20}$ is in the dirty limit.
\begin{figure}
\includegraphics[width=1.0\columnwidth]{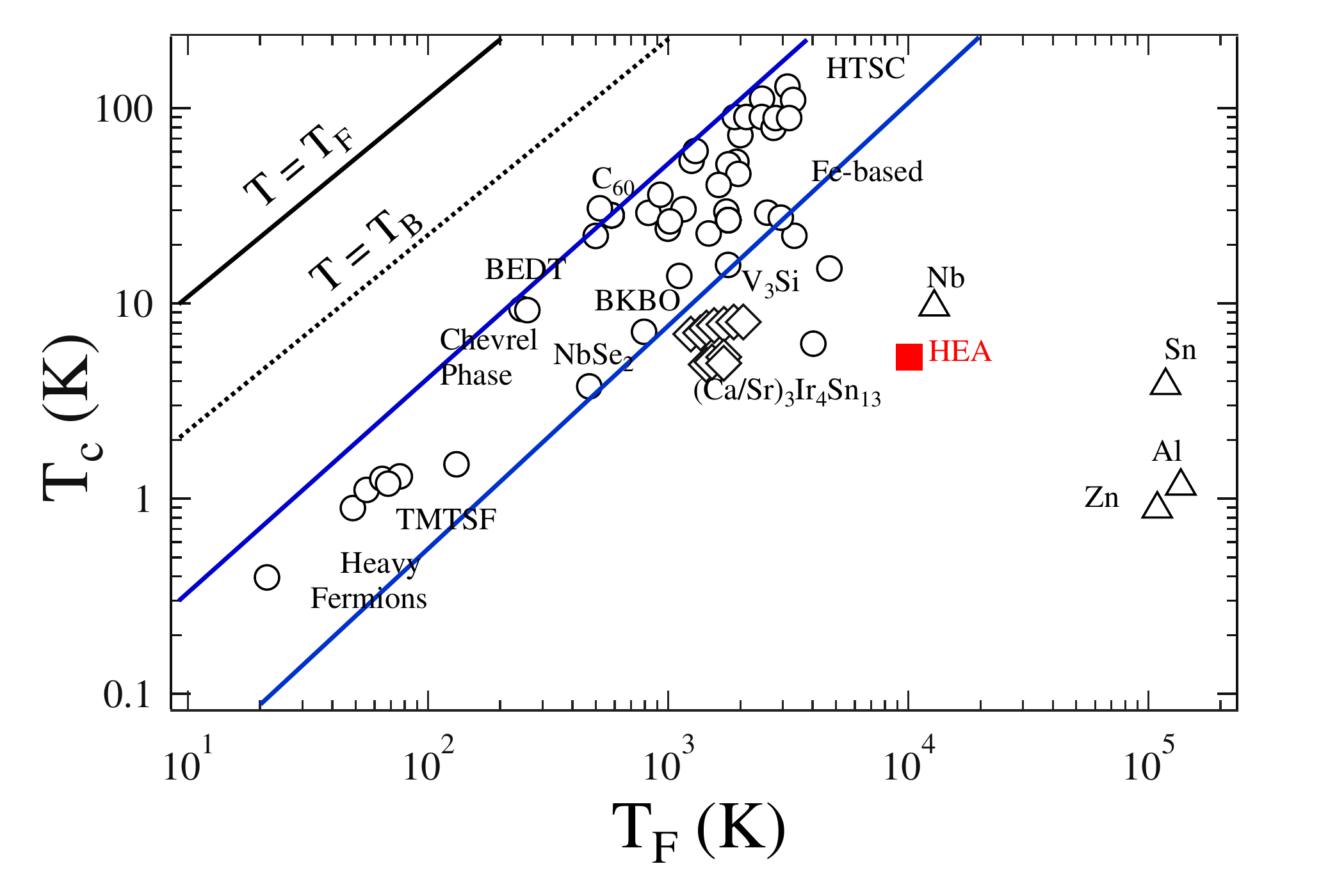}
\caption{\label{Fig7:up} The Uemura plot showing the superconducting transition temperature, $T_{c}$ vs the effective Fermi temperature, $T_{F}$, where Nb$_{21}$Re$_{16}$Zr$_{20}$Hf$_{23}$Ti$_{20}$ (HEA) is shown as a solid red square. Other data points plotted between the blue solid lines are the different families of unconventional superconductors \cite{KKC,RKH}.} 
\end{figure}
Using the estimated value of n in Eq. \ref{eqn6:tf} we have calculated $T_{F}$ = 10091 K. This places Nb$_{21}$Re$_{16}$Zr$_{20}$Hf$_{23}$Ti$_{20}$ away from the unconventional superconductors as shown by a solid red square in Fig. \ref{Fig7:up}, where blue solid lines represent the band of unconventional superconductors.\\

\section{Conclusion}

A single phase polycrystalline HEA material with composition Nb$_{21}$Re$_{16}$Zr$_{20}$Hf$_{23}$Ti$_{20}$ (from EDX analysis and close to the nominal equimolar composition) is prepared by single arc meting furnace. X - ray diffraction pattern indicates that the alloy is arranged on a simple bcc crystal lattice with lattice parameter a = 3.38 \text{\AA}. Transport, magnetization, and thermodynamic measurements reveal that this new HEA is a type-II superconductor with the bulk superconducting transition at $T_{c}$ $\approx$ 5.3 K. This is in fact the first reported bcc structured superconducting HEA having equimolar ratio. In comparison, the bcc structured equimolar Nb-Ta-Hf-Zr-Ti HEA ([TaNb]$_{0.67}$[HfZrTi]$_{0.33}$ shows superconductivity at $T_{c}$ $\approx$ 7.3 K \cite{11}) does not show superconductivity. The obtained value of the upper critical field $H_{c2}$(0)($\approx$ 8.88 T) is close to the Pauli limiting field, and could be attributed to the enhanced SOC due to the introduction of Re in the structure. However, further investigations are required to explore the pairing mechanism and the role of SOC in the sample. Low temperature specific heat measurement indicates that the investigated HEA is close to a BCS-type phonon mediated moderately coupled superconductor.

\section{Acknowledgments}

R.~P.~S.\ acknowledges Science and Engineering Research Board, Government of India for the Young Scientist Grant
No. YSS/2015/001799. S.~M.\ acknowledges Science and Engineering Research Board, Government of India for the NPDF Fellowship (PDF/2016/000348).

\section{Contributions}

S.~M.\ and M.~V.\ equally contributed this work.

\end{document}